\documentstyle[aps,prb,twocolumn]{revtex}
\begin{document}
\draft
\title{Electromagnetic Response of a $k_x\pm ik_y$ Superconductor: 
Effect of Order Parameter Collective Modes}
\author{S. Higashitani and K. Nagai}
\address{Faculty of Integrated Arts and Sciences, Hiroshima University, Kagamiyama 1-7-1, Higashi-hiroshima 739-8521, Japan}
\date{\today}
\maketitle
\begin{abstract}
Effects of order parameter collective modes on electromagnetic 
response are studied for a clean spin-triplet superconductor with 
$k_x\pm ik_y$ orbital symmetry, which has been proposed as a 
candidate pairing symmetry for Sr$_2$RuO$_4$. 
It is shown that the $k_x \pm ik_y$ superconductor has 
characteristic massive collective modes analogous to the clapping mode in 
the A-phase of superfluid $^3$He. We discuss the contribution from the 
collective modes to ultrasound attenuation and electromagnetic absorption. 
We show that in the electromagnetic absorption spectrum the clapping mode 
gives rise to a resonance peak well below the pair breaking frequency, while 
the ultrasound attenuation is hardly influenced by the collective excitations. 
\end{abstract}
\pacs{74.25.Nf, 74.40.+k}

Superconducting and superfluid properties of unconventional Cooper pairing 
states continue to be of interest in the research of strongly correlated 
fermion systems such as liquid $^3$He, heavy fermion compounds, 
high-$T_c$ superconductors and the more recently discovered superconductor 
Sr$_2$RuO$_4$.\cite{Maeno} The spin-triplet $k_x\pm ik_y$-wave state 
is an example of the unconventional pairing states and is known to describe 
the superfluidity of $^3$He-A. Two-dimensional (2D) version of 
this state has recently attracted much 
attention as a candidate pairing state for Sr$_2$RuO$_4$. 

The possibility of the spin-triplet $p$-wave pairing in Sr$_2$RuO$_4$ was 
first suggested by Rice and Sigrist \cite{RS} from 
the analogy to $^3$He. There now exist considerable 
experimental evidences supporting the unconventional spin-triplet 
superconductivity in Sr$_2$RuO$_4$. The absence of the Hebel-Slichter peak 
in $1/T_1$ \cite{IshidaI} and substantial reduction in $T_c$ by non-magnetic 
impurities \cite{Mac} indicate that Sr$_2$RuO$_4$ is at least not 
a conventional $s$-wave superconductor. The $\mu$SR experiments \cite{Luke} 
suggest a pairing state with broken time reversal symmetry. 
More strong evidence of the triplet pairing has been found by $^{17}$O NMR 
measurements \cite{IshidaII} in which the temperature-independent Knight shift 
is observed for magnetic fields parallel to the RuO$_2$ plane. 
Possible pairing states in Sr$_2$RuO$_4$ have been classified from  
a group theoretical point of view.\cite{RS} 
It was shown that the quasi-2D electronic structure of 
Sr$_2$RuO$_4$ leads to five possible $p$-wave states stabilized by the 
absence of gap nodes. Among these states, the pairing state compatible 
with all the above experiments is the $k_x\pm ik_y$ state which is 
the 2D analog of $^3$He-A.

Study of the electromagnetic (EM) response provides valuable 
information on the properties of unconventional superconductors. 
Dynamical properties are, in particular, intriguing because they 
depend not only on the equilibrium gap structure but also on the 
collective excitations of the order parameter.\cite{HN,Hir} 
As is known in the study of superfluid $^3$He, internal degrees of 
freedom of the order parameter give rise to the order parameter 
collective modes (OPCM's) specific to a given pairing symmetry. 
In this paper, we discuss, within the collisionless regime, 
how the OPCM's in the $k_x\pm ik_y$ superconductor contribute 
to the EM response. 

The OPCM's in the 2D $k_x\pm ik_y$ state has been discussed 
by Tewordt.\cite{TeI,TeII} As in the case of $^3$He-A, there exists the 
clapping mode, which is a characteristic massive collective mode 
resulting from the orbital degrees of freedom of the $k_x\pm ik_y$-wave 
order parameter. Tewordt showed that the coupling of this mode to external 
fields vanishes in the long wavelength limit ($q \rightarrow 0$) \cite{TeII} 
and so he did not attempt to discuss the observability of the clapping mode. 
The effect of a finite wave vector ${\bf q}$ is, however, important even 
in type II superconductors when we consider the role of OPCM's in the EM 
response. We in fact show in this paper, by calculating a $q$-dependent 
dynamical conductivity, that the clapping mode leads to an EM absorption peak. 
We also discuss the ultrasound attenuation. We show that, although ultrasound 
experiments have played a key role in the study of the OPCM's in $^3$He, 
the ultrasound cannot be a good probe of the OPCM's in metals such as 
Sr$_2$RuO$_4$. This is because the sound velocity in ordinary metals 
is much smaller than the Fermi velocity $v_{\rm F}$, in contrast to $^3$He.

The 2D $k_x \pm ik_y$ state is defined on a cylindrical 
Fermi surface and the orbital structure of the order parameter 
is expressed in terms of two basis functions ${\hat k}_x = \cos\theta$ and 
${\hat k}_y=\sin\theta$ specifying the direction of the Fermi momentum 
${\bf k}_{\rm F}$. The matrix order parameter is given by 
\begin{equation}
\label{op}
{\hat \Delta}_{\bf k} 
= \Delta ({\hat k}_x \pm i{\hat k}_y) \sigma_z i \sigma_y
= \Delta e^{\pm i\theta}\sigma_z i \sigma_y, 
\end{equation}
where $\sigma_i$'s are Pauli matrices. 
The gap $({\hat\Delta}_{\bf k}{\hat\Delta}_{\bf k}^\dagger)^{1/2} = \Delta$ 
of the two unitary states is independent of ${\bf k}$ and 
the Bogoliubov-quasiparticle energy $E_{\bf k} = 
(\xi_{\bf k}^2+{\hat\Delta}_{\bf k}{\hat\Delta}_{\bf k}^\dagger)^{1/2}$ 
is the same as that of the $s$-wave state. 

The collective excitations of the order parameter are described as 
oscillations of order parameter fluctuations from the equilibrium state.
In the $k_x\pm ik_y$ state, one of the two degenerate states is realized 
in equilibrium. We choose the $k_x + ik_y$ state as the equilibrium 
state and consider an order parameter fluctuation of a plane wave form 
$\delta{\hat\Delta}_{\bf k}({\bf q},\omega)
e^{i({\bf q}\cdot{\bf r}-\omega t)}$. 
Such fluctuation can be excited by space-time dependent EM fields, 
a scalar potential 
$\phi({\bf q},\omega)e^{i({\bf q}\cdot{\bf r}-\omega t)}$ 
and a vector potential 
${\bf A}({\bf q},\omega)e^{i({\bf q}\cdot{\bf r}-\omega t)}$. 
The order parameter fluctuation is expressed using the two bases 
$e^{\pm i\theta}= {\hat k}_x \pm i{\hat k}_y$ as
\begin{equation}
\label{opf}
\delta{\hat \Delta}_{\bf k}({\bf q},\omega) 
= [D_+({\bf q},\omega)e^{i\theta}+D_-({\bf q},\omega)e^{-i\theta}]
\sigma_z i\sigma_y, 
\end{equation}
where $D_+$ and $D_-$ are variables to be determined by the self-consistency 
equation. Since $D_+$ and $D_-$ are complex variables, there are four 
degrees of freedom of the order parameter fluctuations in the representation 
of (\ref{opf}). 

A well-established way to study the dynamical properties including 
the fluctuations of the order parameter is to introduce matrix 
distribution functions in spin space, $n_{\bf k}({\bf q},\omega)$ 
and $f_{\bf k}({\bf q},\omega)$ with matrix elements 
$(n_{\bf k})_{\alpha\beta}=\int dt\,e^{-i\omega t} 
\langle a_{{\bf k}_-\beta}^{\dagger}a_{{\bf k}_+\alpha}^{}\rangle$ and 
$(f_{\bf k})_{\alpha\beta}=\int dt\,e^{-i\omega t} 
\langle a_{-{\bf k}_-\beta}^{}a_{{\bf k}_+\alpha}^{}\rangle$ 
(${\bf k}_\pm = {\bf k}\pm{\bf q}/2$). 
The deviation $\delta f_{\bf k}$ of $f_{\bf k}$ from its equilibrium value 
determines the order parameter fluctuation via the self-consistency equation:
\begin{equation}
\label{seq}
\delta{\hat \Delta}_{\bf k}({\bf q},\omega)
=\sum_{{\bf k}'}v_{{\bf k},{\bf k}'} \delta f_{{\bf k}'}({\bf q},\omega),
\end{equation}
where $v_{{\bf k},{\bf k}'}=-2g_1{\hat{\bf k}}\cdot{\hat{\bf k}}'$ 
is the pairing interaction. The deviation $\delta n_{\bf k}$ is related to 
the charge current density as
\begin{equation}
\label{jeq}
{\bf J}({\bf q},\omega) = ev_{\rm F}\sum_{{\bf k}}{\hat {\bf k}}{\rm Tr}\,
\delta n_{\bf k}({\bf q},\omega)-\frac{ne^2}{mc}{\bf A}({\bf q},\omega),
\end{equation}
where $n$ is the number density. 
(Note that for the spin-independent perturbations considered here, 
$\delta n_{\bf k}$ is proportional to the unit matrix and thus 
the trace in Eq.\ (\ref{jeq}) gives only a factor 2.) 
The distribution functions have been intensively studied in the context of 
ultrasound attenuation in superfluid $^3$He 
(see, for a review, Ref.\ \onlinecite{sbook}). 
Assuming the particle-hole symmetry, we can write the distribution functions 
integrated over the energy variable $\xi_{\bf k}$ as
\begin{eqnarray}
\label{neq}
\int_{-\infty}^{\infty}d\xi_{\bf k}\delta n_{\bf k}
&=&(\delta\epsilon_0+\delta\epsilon_1)\eta/(\omega-\eta)\cr
&-&2{\cal F}{\hat\Delta}_{\bf k}{\hat\Delta}_{\bf k}^\dagger
(\omega\delta\epsilon_0+\eta\delta\epsilon_1)/(\omega-\eta)\cr
&+&\frac{{\cal F}}{2}(\omega+\eta)(\delta{\hat\Delta}_{\bf k}{\hat\Delta}_{\bf k}^\dagger-{\hat\Delta}_{\bf k}\delta{\hat\Delta}_{\bf k}^\dagger),\\
\label{feq}
\int_{-\infty}^{\infty}d\xi_{\bf k}\delta f_{{\bf k}}
&=&\int_{-\varepsilon_c}^{\varepsilon_c}d\xi_{\bf k}\frac{\Theta_{\bf k}}{E_{\bf k}}\delta{\hat \Delta}_{\bf k}\cr 
&-&{\cal F}\Bigl[\frac{1}{2}(\omega^2-2{\hat \Delta}_{\bf k}{\hat \Delta}_{\bf k}^\dagger-\eta^2)\delta{\hat \Delta}_{\bf k}\cr
&&\quad-{\hat \Delta}_{\bf k}\delta{\hat \Delta}_{\bf k}^\dagger{\hat \Delta}_{\bf k}-(\omega\delta\epsilon_0+\eta\delta\epsilon_1){\hat \Delta}_{\bf k}\Bigr],
\end{eqnarray}
where $\delta\epsilon_0 = e\phi({\bf q},\omega)$ and $\delta\epsilon_1 = 
-(e/mc){\bf k}_{\rm F}\cdot{\bf A}({\bf q},\omega)$ represent the perturbation 
energies, $\eta = {\bf v}_{\rm F}\cdot{\bf q}$, 
$\Theta_{\bf k} = -(1/2)\tanh(E_{\bf k}/2T)$, $\varepsilon_c$ 
is the usual cutoff energy and 
\begin{eqnarray}
\label{fdef}
{\cal F}=\int_{-\infty}^{\infty}
\frac{d\xi_{\bf k}}{2E_{{\bf k}_+}E_{{\bf k}_-}}
&&\biggl[\frac{(E_{{\bf k}_+}+E_{{\bf k}_-})
(\Theta_{{\bf k}_-}+\Theta_{{\bf k}_+})}
{\omega^2-(E_{{\bf k}_+}+E_{{\bf k}_-})^2}\cr
&&+\frac{(E_{{\bf k}_+}-E_{{\bf k}_-})
(\Theta_{{\bf k}_-}-\Theta_{{\bf k}_+})}
{\omega^2-(E_{{\bf k}_+}-E_{{\bf k}_-})^2}\biggr].
\end{eqnarray}
Here we have dropped the arguments ${\bf q}$ and $\omega$ for brevity
($\delta{\hat\Delta}^\dagger = \delta{\hat\Delta}^\dagger(-{\bf q},-\omega)$).
In Eqs.\ (\ref{neq}) and (\ref{feq}), small corrections of order 
$q/k_{\rm F}$ have been neglected and accordingly ${\hat\Delta}_{{\bf k}_\pm}$ 
has been put as ${\hat\Delta}_{{\bf k}}$.\cite{Goryo} 

Now we discuss the OPCM's in the $k_x+ik_y$ state. 
It is convenient to introduce the following new variables:
\begin{eqnarray}
&&\frac{1}{2}[D_+({\bf q}, \omega) \pm D_+^*(-{\bf q}, -\omega)]
= \cases{D_+' \cr D_+''},\\
&&\frac{1}{2}[D_-({\bf q}, \omega)e^{-2i\theta_q} 
\pm D_-^*(-{\bf q}, -\omega)e^{2i\theta_q}]= \cases{D_-' \cr D_-''},
\end{eqnarray}
where $e^{i\theta_q}={\hat q}_x+{\rm i}{\hat q}_y$. 
Using Eq.\ (\ref{feq}), the self-consistency equation (\ref{seq}) 
can be written in terms of these variables as 
\begin{eqnarray}
\label{seqone}
M_1 \pmatrix{D_+' \cr D_-'} &=&
\pmatrix{0 \cr 2i\Delta\langle {\cal F}(\omega\delta\epsilon_0+\eta\delta\epsilon_1)\sin 2\vartheta\rangle_{\bf k}},\\
\label{seqtwo}
M_2 \pmatrix{D_+'' \cr D_-''} &=&
\pmatrix{2\Delta\langle{\cal F}(\omega\delta\epsilon_0+\eta\delta\epsilon_1)\rangle_{\bf k} \cr 2\Delta\langle {\cal F}(\omega\delta\epsilon_0+\eta\delta\epsilon_1)\cos 2\vartheta\rangle_{\bf k}},
\end{eqnarray}
where $\vartheta$ is the angle between ${\bf k}$ and ${\bf q}$, 
$\langle\cdots\rangle_{\bf k}$ denotes the angle average over the Fermi 
surface and $M_{1,2}$ are $2\times 2$ matrices with the following elements:
\begin{eqnarray}
(M_1)_{11} &=& \langle{\cal F}(\omega^2-4\Delta^2-\eta^2)\rangle_{\bf k},\cr
(M_1)_{22} &=& \langle{\cal F}[\omega^2-2\Delta^2(1+\cos 4\vartheta)-\eta^2]\rangle_{\bf k},\cr
(M_1)_{12} &=& (M_1)_{21}=\langle{\cal F}(\omega^2-4\Delta^2-\eta^2)\cos 2\vartheta\rangle_{\bf k},\cr
(M_2)_{11} &=& \langle{\cal F}(\omega^2-\eta^2)\rangle_{\bf k},\cr
(M_2)_{22} &=& \langle{\cal F}[\omega^2-2\Delta^2(1-\cos 4\vartheta)-\eta^2]\rangle_{\bf k},\cr
(M_2)_{12} &=& (M_2)_{21}=\langle{\cal F}(\omega^2-\eta^2)\cos 2\vartheta\rangle_{\bf k}.
\end{eqnarray}
Note that the variables $D'$ and $D''$ are decoupled from each other. 

The OPCM's correspond to the eigen modes in the absence of 
external fields. The eigen frequencies $\omega$ are therefore 
determined by $\det M_{1,2}=0$.
In the long wavelength limit, $q \rightarrow 0$, it is easy to obtain 
the eigen frequencies. In this limit, since the off-diagonal 
elements of $M_{1,2}$ may be neglected, all the modes decouple. 
The results for $q\rightarrow 0$ are summarized in Table \ref{taI}. 
The $D_+'$-mode has the eigen frequency just at the pair breaking edge, 
$\omega = 2\Delta$. The variable $D_+''$ represents a mode corresponding to 
the phase fluctuation of the order parameter, namely, the so-called 
Anderson-Bogoliubov mode. This is a gapless mode with sound-like dispersion 
$\omega=(v_{\rm F}/\sqrt{2})q$ but is replaced by a plasmon when the Coulomb 
interaction between electrons is taken into account.\cite{HN} 
The remaining two modes, $D_-$-modes, have the same character 
as the clapping mode in superfluid $^3$He-A.\cite{sbook,Wolf} 
The eigen frequency of the clapping modes in the 2D system 
is $\omega = \sqrt{2}\Delta$ and is below the pair breaking energy $2\Delta$. 
At finite $q$, the dispersion relation of 
the clapping modes up to the order $q^2$ is given by
\begin{equation}
\omega^2_{\rm cl}(q) = 2\Delta^2 + \frac{1}{2}(v_{\rm F}q)^2. 
\end{equation}

The coupling of the OPCM's to the external fields is determined by 
the right-hand side of Eqs.\ (\ref{seqone}) and (\ref{seqtwo}). We see that 
the $D'_\pm$-modes and the $D''_\pm$-modes couple to the transverse 
field and the longitudinal field, respectively.

The presence of the coupling between the $D''_\pm$-modes and the longitudinal 
field means that these modes can be excited by a longitudinal phonon field. 
Since the collective excitations couple to the density fluctuation 
$\delta\rho({\bf q},\omega)=\sum_{\bf k}{\rm Tr}\,\delta n_{\bf k}$ 
via the last term in Eq.\ (\ref{neq}), there is a possibility 
that the clapping mode ($D_-''$-mode) is observed by ultrasound measurements. 
We have estimated the ultrasound attenuation coefficient by assuming 
typical conditions $\omega \ll v_{\rm F}q \ll \Delta$ for ultrasound 
measurements in ordinary metals and also in Sr$_2$RuO$_4$. 
(In Sr$_2$RuO$_4$, the sound velocity is $\sim 10^5$ cm/s \cite{Matsui}, 
the Fermi velocity is $\sim 10^7$ cm/s \cite{Oguchi,deH} 
and $\Delta \sim 1$ K.) 
The attenuation coefficient  from quasiparticle excitations 
coincides with that for the $s$-wave superconductor because of 
the isotropic gap. The contribution from the collective excitations 
can be obtained using Eqs.\ (\ref{neq}) and (\ref{seqtwo}). 
Taking into account the above conditions and considering the limit 
$\omega \rightarrow 0$, one can find, after straightforward algebra, 
that the collective excitations give only a small correction of order 
$(\omega/v_{\rm F}q)^2 \sim 10^{-4}$ to the quasiparticle contribution. 
Consequently, the ratio of the superconducting to normal attenuation 
coefficient in the low-frequency limit $\omega \rightarrow 0$ is given by 
the well-known BCS result $\alpha_s/\alpha_n = 2/(e^{\Delta/T}+1)$. 
Thus sound waves are not suitable probe for the OPCM. 
It is to be noted that, in contrast to ordinary metals, 
ultrasound propagation in superfluid $^3$He is in the high frequency regime 
such that $\omega \sim \Delta \gg v_{\rm F}q$ where the OPCM plays 
an important role. This is a reason why sound waves can be a good probe 
of the OPCM's in superfluid $^3$He. 

Now we consider the transverse response of a $k_x+ik_y$ superconductor. 
The transverse response is conveniently described by the complex 
conductivity $\sigma_t({\bf q},\omega)$ for the transverse electric field 
$E_t({\bf q},\omega)=(i\omega/c)A_t({\bf q},\omega)$. 
The real part $\sigma_1$ of the complex conductivity 
$\sigma_t = \sigma_1 + i\sigma_2$ determines the absorption of EM waves. 
We can readily obtain $\sigma_t$ by calculating 
the transverse current density using the above formulation. 
The conductivity consists of three characteristic parts, i.e., 
$\sigma_t = \sigma_t^{\rm qp}+\sigma_t^{\rm cm}+\sigma_t^{\rm dia}$; 
the first and the second terms are contributions from quasiparticle 
excitations and from collective modes, respectively, and the last term 
$\sigma_t^{\rm dia} = ine^2/m\omega$ arises from the diamagnetic current 
$(-ne^2/mc)A_t$. The quasiparticle contribution $\sigma_t^{\rm qp}$ is 
the same as the BCS result for the $s$-wave superconductor:
\begin{equation}
\label{sqp}
\sigma_t^{\rm qp}({\bf q},\omega)
= \frac{2ine^2}{m\omega}\langle\frac{{\hat k}_t^2\eta^2}{\omega^2-\eta^2}(1-2\Delta^2{\cal F})\rangle_{\bf k},
\end{equation}
where ${\hat k}_t = \sin\vartheta$ is the transverse component of the unit 
vector ${\hat {\bf k}}$. 
The collective-mode contribution $\sigma^{\rm cm}_t$ comes only from 
the $D_-'$-clapping mode and is obtained as
\begin{eqnarray}
\label{scm}
\sigma_t^{\rm cm}&&({\bf q},\omega)
=-2iev_{\rm F}\Delta\sum_{\bf k}{\hat k}_t\eta{\cal F}
\sin 2\vartheta D_-'/E_t\cr
=&& \frac{4ine^2}{m\omega}\frac{\Delta^2\langle{\cal F}(\omega^2-4\Delta^2-\eta^2)\rangle_{\bf k}\langle{\cal F}\eta{\hat k}_t\sin 2\vartheta\rangle_{\bf k}^2
}{\det M_1},
\end{eqnarray}
where the last line is derived using $D_-'$ determined by Eq.\ (\ref{seqone}). 
Note that Eq.\ (\ref{scm}) includes $\det M_1$ in the denominator. 
This implies that a resonance EM absorption due to 
the clapping mode occurs at the eigen frequency. 
To see this explicitly, it is useful to estimate the real 
part $\sigma_1$ by assuming the long wavelength condition such that 
$v_{\rm F}q \ll \Delta$. This assumption is justified for type II 
superconductors, since the penetration depth $\lambda$ is large 
compared with the coherence length $\xi_0$ and the important values 
of $q$ in the Fourier integral of the fields penetrating into 
the superconductor are $\lesssim \lambda^{-1}$. In what follows, 
we discuss the $\omega$-dependence of $\sigma_1$ in this case. 

Since we are interested in high frequencies 
$\omega \sim \Delta$, we may expand the conductivity 
in terms of $q$. Expanding Eqs.\ (\ref{sqp}) and (\ref{scm}) up to 
the order $q^2$ and adding the results, we obtain 
\begin{equation}
\label{stlwll}
\sigma_t({\bf q},\omega) = 
\frac{ine^2v_{\rm F}^2q^2}{4m\omega^3}\left(1-{\cal F}_0\frac{\omega^2-4\Delta^2}{\omega^2-\omega_{\rm cl}^2}\right)+\sigma_t^{\rm dia},
\end{equation}
where ${\cal F}_0 = \Delta^2{\cal F}({\bf q}=0,\omega)$. 
Equation (\ref{stlwll}) clearly shows that $\sigma_t$ has a pole at 
the eigen frequency of the clapping mode, yielding a delta function peak 
in the real part $\sigma_1$. The diamagnetic term $\sigma_t^{\rm dia}$ 
in Eq. (\ref{stlwll}) is pure imaginary except at $\omega = 0$, so that only 
the first term contributes to $\sigma_1$ at the high frequencies of interest. 

We  briefly discuss $\sigma_1$ in the low frequency region 
$\omega \ll \Delta$ where the OPCM does not come into play. 
There are two contributions to $\sigma_1$ at the low frequencies. 
One is a delta function at $\omega=0$, associated with superfluid motion 
excited by EM fields. 
The other is caused by thermally excited quasiparticles. 
The energy conservation $\omega = |E_{{\bf k}_+}-E_{{\bf k}_-}|$ for 
the corresponding quasiparticle-scattering process requires that this 
contribution is restricted in the region $\omega < v_{\rm F}q \ll \Delta$. 
We note that in the limit $v_{\rm F}q/\Delta \rightarrow 0$, 
the conductivity sum rule 
$\int_0^\infty d\omega \sigma_1({\bf q},\omega)=\pi ne^2/2m$ 
is satisfied only by the two contributions. 
As can be shown from Eq.\ (\ref{sqp}), 
in the limit $v_{\rm F}q/\Delta \rightarrow 0$, 
the $\omega$-integration yields $\pi n_se^2/2m$ for the $\delta(\omega)$ term 
and $\pi n_ne^2/2m$ for the contribution from thermally excited 
quasiparticles, where $n_s$ is the superfluid density and $n_n = n-n_s$ is 
the normal-fluid density. 
This explains why $\sigma_1$ at $\omega \sim \Delta$, given by 
Eq.\ (\ref{stlwll}), is proportional to $q^2$.

\begin{figure}[h]
\special{epsfile=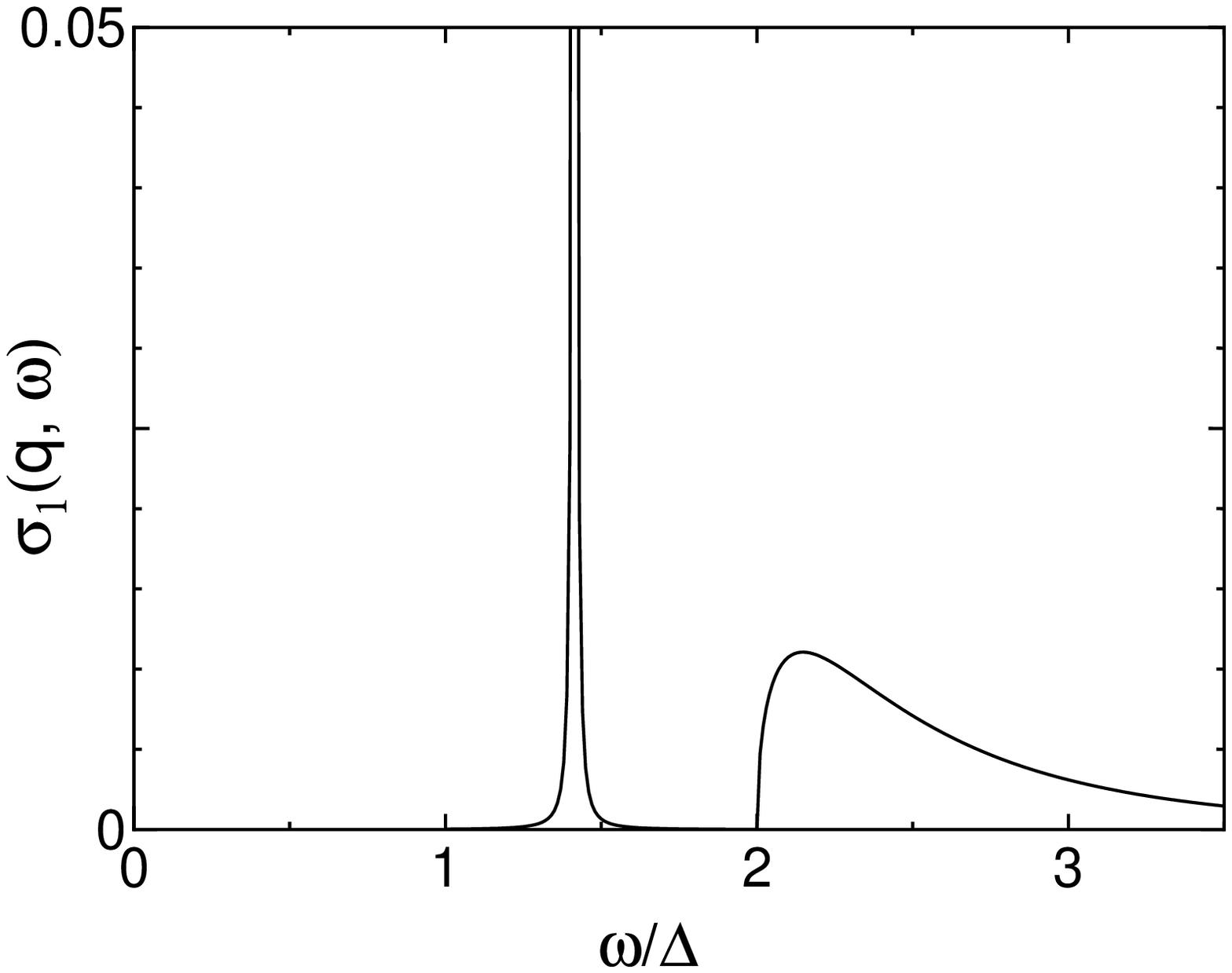 hsize=8cm}
\vspace{7cm}
\caption{Real part $\sigma_1({\bf q},\omega)$ of the complex 
conductivity at $T=0$ as a function of reduced frequency $\omega/\Delta$.
The ordinate is scaled by $ne^2v_{\rm F}^2q^2/m\Delta^3$. 
A delta function peak at the eigen frequency 
of the clapping mode, $\omega/\Delta = \sqrt{2}$, is plotted 
as a Lorentzian with a tiny width.}
\label{sw}
\end{figure}

Equation (\ref{stlwll}) depends on temperature $T$ through ${\cal F}_0$ 
and $\Delta$. At sufficiently low temperatures, the hyperbolic tangent 
function, $\tanh(E_{\bf k}/2T)$, in ${\cal F}_0$ may be replaced by unity; 
then Eq.\ (\ref{stlwll}) is reduced to the result for $T=0$. 
Figure \ref{sw} shows the $\omega$-dependence of the real part 
$\sigma_1$ at $T=0$ obtained from Eq.\ (\ref{stlwll}). 
The structure above $\omega=2\Delta$ is caused by pair-breaking processes. 
Below the pair-breaking edge, $\omega < 2\Delta$, 
there is a delta function peak at $\omega_{\rm cl} = \sqrt{2}\Delta$, 
associated with the absorption of the clapping mode. This collective-mode 
contribution is explicitly given by $\sigma_1^{\rm cm}({\bf q},\omega) = 
(\pi^2ne^2v_{\rm F}^2q^2/64m\Delta^2)\delta(\omega-\omega_{\rm cl})$.

Hirschfeld {\it et al}.\cite{Hir} have studied the power absorption 
from the OPCM in the Balian-Werthamer state (in a 3D system). 
The power absorption $P(\omega)$ is given as the dissipation of 
the field energy which appears as Joule heat. To consider $P(\omega)$, 
we need to take into account the presence of the vacuum-metal interface. 
They estimated $P(\omega)$ by assuming the specular surface boundary condition 
and by neglecting the surface scattering effect causing 
the pair breaking near the surface. In the 2D $k_x\pm ik_y$ 
state considered here, the same calculation is repeated by using 
Eq.\ (\ref{stlwll}) and considering the EM wave injected 
along the 2D plane. The resulting collective-mode contribution 
to $P(\omega)$ has the following $\omega$-dependence: $P^{\rm cm}(\omega) = 0$ 
below $\omega = \omega_{\rm cl}(0)$. Above the threshold, 
$\omega > \omega_{\rm cl}(0)$, $P^{\rm cm}(\omega)$ increases rapidly with 
$\omega$ and then has a peak structure with a finite width 
$\sim (\xi_0/\lambda_L)\Delta$ ($\lambda_L = (mc^2/4\pi ne^2)^{1/2}$, 
the London penetration depth); this $\omega$-dependence is descried by 
$P^{\rm cm}(\omega) \sim [\omega^2 - \omega_{\rm cl}^2(0)]^{1/2}
/[\omega^2-\omega_{\rm cl}^2(0)+c_1^2\lambda_L^{-2}]^2$, 
where $c_1 = v_{\rm F}/\sqrt{2}$ is the velocity determining 
the dispersion of the clapping mode.
In the power absorption spectrum, the OPCM gives rise to 
not a delta function peak but a broadened peak. This is 
a consequence of taking into account the dispersion of 
$\omega_{\rm cl}$ in the denominator of Eq.\ (\ref{stlwll}). 
In actual metals, impurity scattering also brings about the broadening. 
The power absorption spectrum, however, has a definite peak structure due to 
the collective excitation if $\xi_0$ is small enough compared with $\lambda_L$ 
and at least the superconductor is clean. This demonstrates that the clapping 
mode in the $k_x\pm ik_y$-wave type II superconductor can be 
observed by EM absorption measurements.


\begin{table}[h]
\caption{Order parameter collective modes of the $k_x + ik_y$ state 
in the long wavelength limit $q \rightarrow 0$.}
\label{taI}
\begin{tabular}{ccc} 
Variable & Eigen frequency & Mode \\
\hline
$D_+'$ & $2\Delta$ &  \\
$D_+''$ & $0$ & Anderson-Bogoliubov mode  \\
$D_-'$, $D_-''$ & $\sqrt{2}\Delta$ & Clapping mode  \\
\end{tabular} 
\end{table}
\end{document}